\documentclass[aip,reprint]{revtex4-1}
\usepackage{graphicx}

\draft 

\begin{document}


\title{Dual-Comb Photothermal Microscopy}

\author{Peter $\text{Chang}^{*,\#}$}
\affiliation{Electrical, Computer and Energy Engineering, University of Colorado, Boulder, CO, USA}
\affiliation{Department of Physics, University of Colorado, Boulder, CO, USA}

\author{Ragib $\text{Ishrak}^*$}
\affiliation{Department of Electrical and Computer Engineering, University of Houston, Houston, TX, USA}

\author{Nazanin Hoghooghi}
\affiliation{Precision Laser Diagnostics Laboratory, University of Colorado, Boulder, CO, USA}

\author{Scott Egbert}
\affiliation{Precision Laser Diagnostics Laboratory, University of Colorado, Boulder, CO, USA}

\author{Gregory B. Rieker}
\affiliation{Precision Laser Diagnostics Laboratory, University of Colorado, Boulder, CO, USA}

\author{Scott A. $\text{Diddams}^{\dagger}$}
\affiliation{Electrical, Computer and Energy Engineering, University of Colorado, Boulder, CO, USA}
\affiliation{Department of Physics, University of Colorado, Boulder, CO, USA}

\author{Rohith Reddy}
\affiliation{Department of Electrical and Computer Engineering, University of Houston, Houston, TX, USA}

\begin{abstract}
\textit{\textsuperscript{*} these authors contributed equally} \\
\textit{\textsuperscript{$\#$} peter.chang-1@colorado.edu} \\
\textit{\textsuperscript{$\dagger$} scott.diddams@colorado.edu}\\
(Dated: 12 September 2024) \\ \\
%
%
We introduce a new parallelized approach to photothermal microscopy that utilizes mid-infrared dual-comb lasers, enabling simultaneous measurements at hundreds of wavelengths. This technology allows for high-speed, label-free chemical identification with super-resolution infrared imaging, overcoming the limitations of slow, single-wavelength-laser methods.
\end{abstract}

\pacs{}

\maketitle 

Current microscopy approaches struggle to meet the combination of sub-cellular spatial resolution, label-free chemical specificity, and fast acquisition to characterize complex bio-molecular systems in real-time. Addressing this need, we combine rapid and broad bandwidth mid-infrared (MIR) spectroscopy via dual frequency combs with photothermal microscopy to provide label-free chemical identification with super-resolution. By parallelizing photothermal excitation through a dual-comb architecture, we open the possibility of collecting MIR spectra across 1000 $\rm cm^{-1}$ with the spatial resolution provided by a visible probe laser. This new imaging modality can address real-time, in situ sub-cellular imaging of bio-molecular constituents such as amino acids, glucose, RNA, and protein secondary structures required for the quantitative monitoring of spatio-temporal molecular dynamics in complex bio-systems.

\begin{figure}[!ht]
\centering
\includegraphics[width=\linewidth]{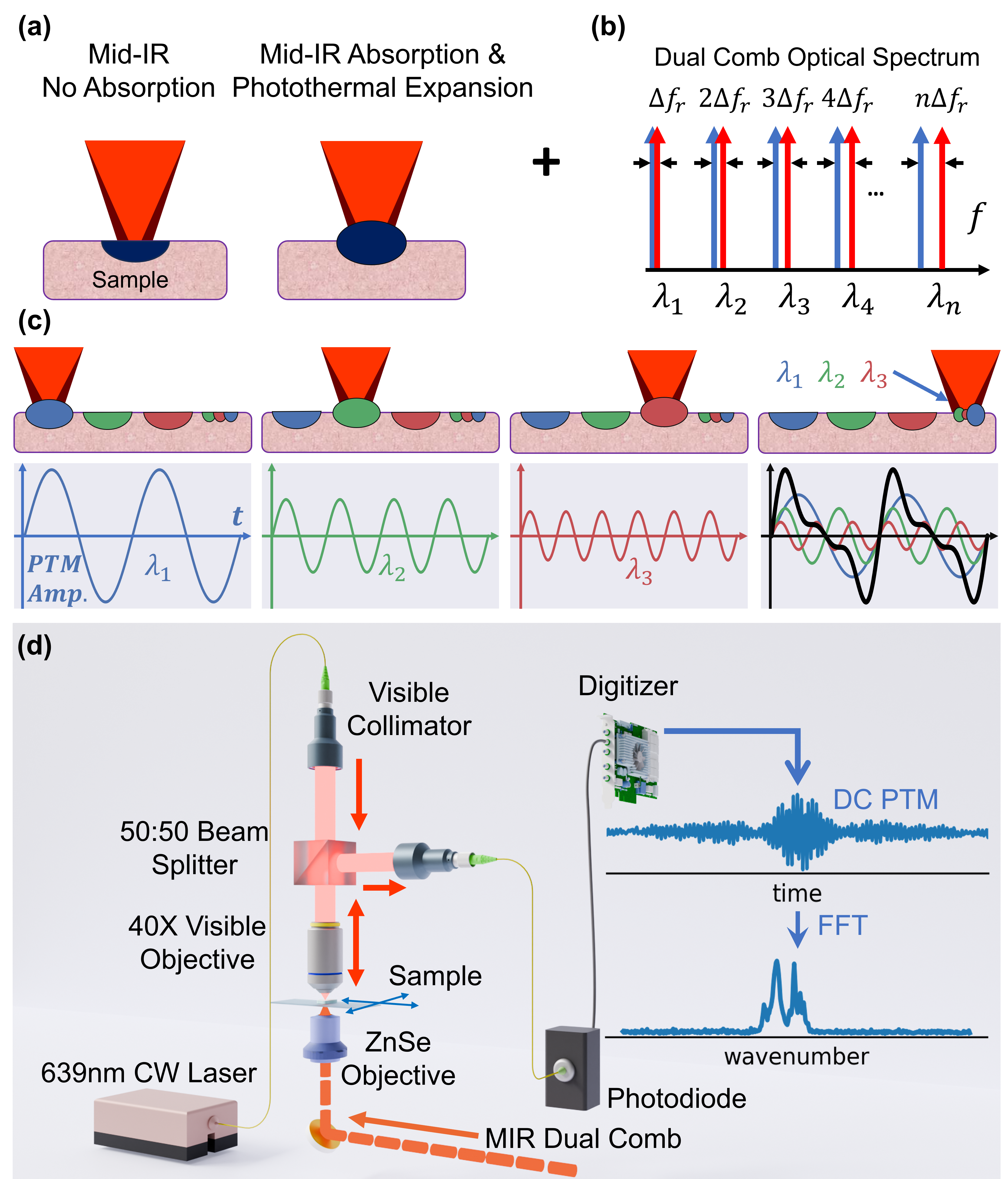}
\caption{Concept of DC-PTM. (a) Absorption of resonant MIR light leads to local photothermal expansion. (b) A dual-comb optical spectrum consists of multiple pairs of comb modes separated by $n\Delta f_r$.  (c) Hyperspectral photothermal response at $\lambda_n$ is measured at absorbing locations that correspond to excitation from different comb mode pairs. (d) Schematic of DC-PTM with counterpropagating MIR dual comb pump and visible light probe.}
\label{fig:setup}
\end{figure}

The concepts central to dual-comb photothermal microscopy (DC-PTM) are illustrated in Figure \ref{fig:setup}.   In conventional photothermal microscopy (Fig. \ref{fig:setup}a), a MIR pump laser is amplitude modulated at $f_m$ and focused onto the sample. A modulated photothermal response is induced if the molecules at a sample location absorb a specific MIR wavelength light. This leads to a photothermal expansion at $f_m$ that can be probed and read out in the backscattered visible probe. With $\lambda_{\rm probe}=0.639~\mu \rm{m}$ and  $\rm{NA}=0.65$, the resolution is $0.5~\mu \rm{m}$, or $10\times$ better than the $4-10~\mu {\rm m}$ resolution of the MIR pump. A challenge limiting the widespread adoption of photothermal microscopy has been its slow data acquisition speed, as each wavelength in the MIR must be scanned sequentially. 

Here, we leverage advances in GHz MIR optical frequency combs and dual-comb spectroscopy (DCS) \cite{hoghooghiBroadband1GHzMidinfrared2022,hoghooghiGHzRepetitionRate2024} to parallelize the MIR photothermal pump. While dual-comb techniques have been applied to gaseous photothermal spectroscopy \cite{wangDualcombPhotothermalSpectroscopy2022}, they have not been used in photothermal imaging. Dual-comb methods have also been employed alongside fluorescence-detected photothermal infrared imaging (F-PTIR) \cite{turnerDualCombMidInfraredSpectromicroscopy2024}. However, F-PTIR often requires labeling, and recent work has not transduced dual-comb interference into the fluorescence signal. 

In our experiments, we use two MIR frequency combs with slightly offset repetition rates ($\Delta f_r$). These combs provide multiple pairs of comb modes that mix to produce modulated MIR pumps at $f_m=n\Delta f_r$. These modulation tones are precisely arranged across the MIR spectrum at wavelengths $\lambda_n$ (see Fig. \ref{fig:setup}b), and create a direct link between MIR spectroscopic signals and the modulation $n\Delta f_r$ (see Fig. \ref{fig:setup}c). By employing the setup shown in Fig. \ref{fig:setup}d, we use a visible-wavelength probe to capture the multi-frequency modulation, which is then demodulated through a Fourier Transform. This process allows us to simultaneously reveal spatially varying MIR chemical fingerprints across the frequency comb bandwidth.

The 1-GHz MIR dual combs are generated by driving intrapulse difference-frequency generation with few-cycle pulses centered at 1550 nm in periodically poled Lithium Niobate (PPLN), operating in the $3-5~\mu{\rm m}$ range, though longer wavelengths are possible \cite{hoghooghiBroadband1GHzMidinfrared2022}. The two MIR combs are combined on a $\mathrm{CaF_2}$ beam splitter, focused onto the sample (\mbox{$\sim$ 10 mW}) with a 0.25 NA ZnSe lens. A $\mathrm{CaF_2}$ window picks off a small portion of the combined combs before the microscope to collect the dual-comb background on an MCT detector, allowing adjustment of repetition rate differences and serving as a high SNR phase-correction template for extended averaging times.


The MIR dual-combs illuminate the sample from below (\mbox{Fig. \ref{fig:setup}}.(d)), while the visible probe is focused from above using a 40x objective of NA 0.65. The multi-wavelength heterodyne of the the mid-infrared combs produces an intensity modulation that is encoded onto the back-scattered visible probe beam via the photothermal effect. This probe light is detected and digitized, and the interferogram train is phase-corrected \cite{yuDigitalErrorCorrection2019} with the parameters calculated from the MIR dual-comb. The averaged interferogram is apodized from \mbox{0.033 $\rm{cm^{-1}}$} to \mbox{5 $\rm{cm^{-1}}$} and Fourier transformed to yield the photothermal spectrum. This spectrum is normalized to the incident mid-infrared power spectral density to obtain the sample's absorption spectrum.

\begin{figure}[!ht]
\centering
\includegraphics[width=\linewidth]{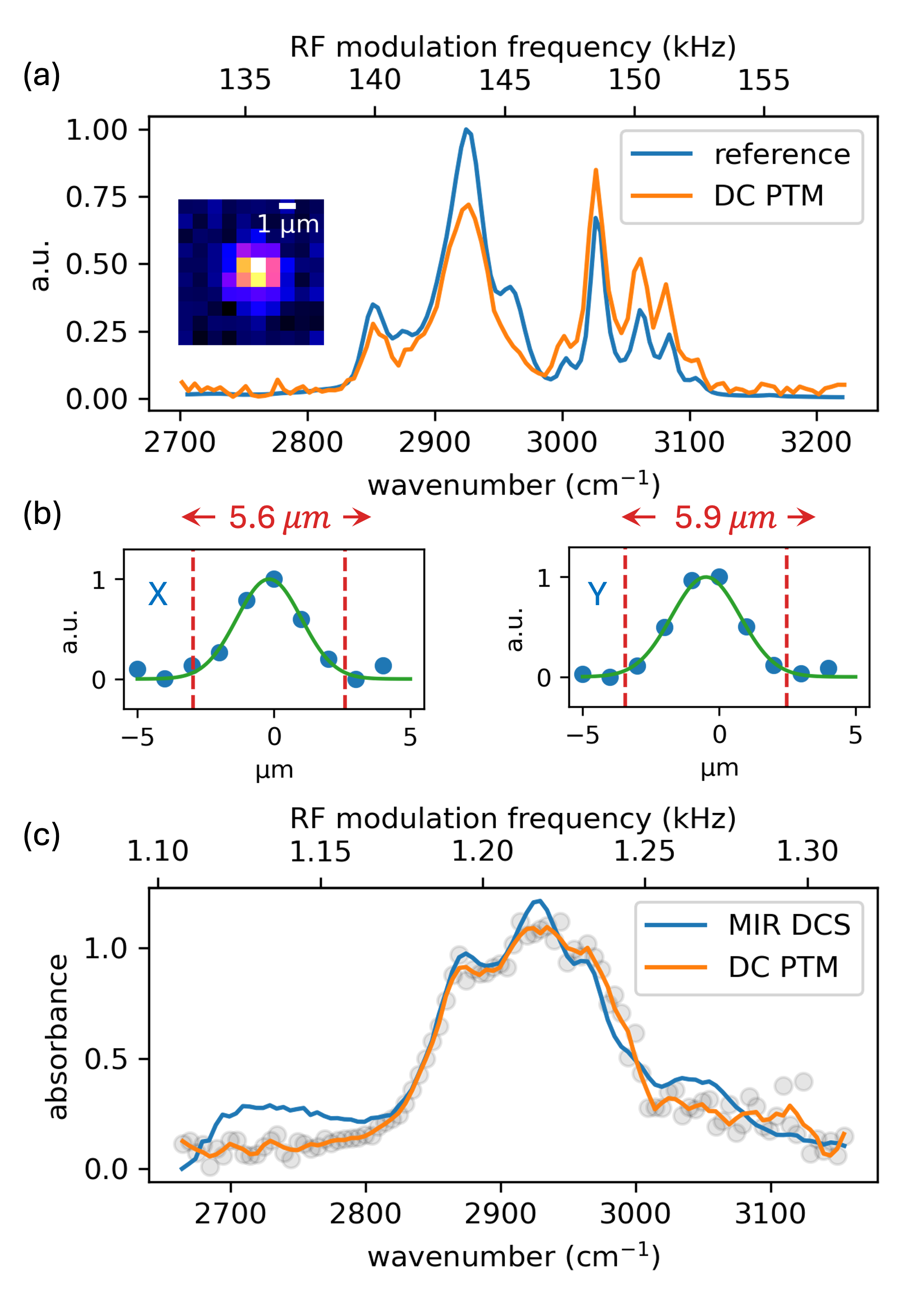}
\caption{Photothermal Dual-Comb Mid-Infrared Spectroscopy: (a) Photothermal dual-comb spectrum of polystyrene (orange) compared to reference data (blue) \cite{myersIARPAPNNLLiquid}. The inset shows a DC-PTM image of a polystyrene bead, (b) Marginals of the image in (a) shown with Gaussian fits (orange) used to estimate the feature size given as the $2\sigma$ span of the curve, (c) DC-PTM spectrum of SU-8 photoresist compared to a MIR dual-comb spectrum (gray). The orange curve shows the raw data (gray) smoothed by a Savitzky–Golay filter.}
\label{fig:results}
\end{figure}

To demonstrate the advantages and potential for super-resolution with DC-PTM we obtain spatio-spectral information of a single \mbox{5 $\mu \rm{ m}$} diameter polystyrene bead and  SU-8 photoresist. In \mbox{Figure \ref{fig:results}} we compare the photothermal spectrum of the bead to a polystyrene reference spectrum \cite{myersIARPAPNNLLiquid}, and the spectrum of the SU-8 sample that is measured by mid-infrared dual-comb. The photothermal spectra compare well to the expected results apart from differences in SNR, which is more noticeable in the case of SU-8. As revealed by the RF axes in \mbox{Fig. \ref{fig:results}. (a, c)}, the SU-8 photothermal response is 100$\times $ slower than that of the polystyrene bead, due to the \mbox{$\sim 12  \mu {\rm m}$} thickness of the sample.


\mbox{Fig. \ref{fig:results}}b shows the integrated marginals of the acquired bead image that are fitted by Gaussian curves. This represents the convolution of the bead's size with the spot-size of the focused probe. Considering the probe \mbox{$\lambda_{\text{vis}}=639$ nm}, the microscope's spatial resolution is limited by the bead size. Nonetheless, the image resolution in \mbox{Fig. \ref{fig:results}. (a)} can be estimated with a Fourier ring correlation \cite{nieuwenhuizenMeasuringImageResolution2013} to be \mbox{$\sim 2.6\; \mu m$}, which is well  below the \mbox{$\sim 7 \; \mu m$} diffraction limit for \mbox{$\lambda_{\text{mir}}=3.5 \; \mu m$}. This demonstrates the super-resolution potential of DC-PTM. 


\smallskip
\noindent\textbf{Funding.} National Science Foundation 2019195, CPRIT RR170075, NIH grant R01DK135870 and R01HL173597 (RR).

\smallskip
\noindent\textbf{Disclosures.} The authors declare no conflicts of interest.

\smallskip
\noindent\textbf{Data Availability Statement.} Data underlying these results are available from the authors upon reasonable request.

\bibliography{main}

\end{document}